\documentclass{PoS}

\title{EPS09 - Global NLO analysis of nuclear PDFs and their uncertainties}

\ShortTitle{EPS09 - Global NLO analysis of nuclear PDFs and their uncertainties}

\author{K. J. Eskola, \speaker{H. Paukkunen} \\

    Department of Physics, University of Jyv\"askyl\"a and 
		Helsinki Institute of Physics, Finland \\

        E-mail: \email{kari.eskola,hannu.paukkunen@phys.jyu.fi}}		

\author{C. A. Salgado\\

     Departamento de F\'\i sica de Part\'\i culas and IGFAE, Universidade de Santiago de 
     Compostela, Spain\\

        E-mail: \email{carlos.salgado@usc.es}}

\abstract{In this talk, we introduce our recently completed next-to-leading order (NLO) global analysis of the nuclear parton distribution functions (nPDFs) called EPS09 --- a higher order successor to the well-known leading-order (LO) analysis EKS98 and also to our previous LO work EPS08. As an extension to similar global analyses carried out by other groups, we complement the data from deep inelastic $l+A$ scattering and Drell-Yan dilepton measurements in p+$A$ collisions by inclusive midrapidity pion production data from d+Au collisions at RHIC, which results in better constrained gluon distributions than before. The most important new ingredient, however, is the detailed error analysis, which employs the Hessian method and which allows us to map out the parameter-space vicinity of the best-fit to a collection of nPDF error sets.
These error sets provide the end-user a way to compute how the PDF-uncertainties will propagate into the cross sections of his/her interest. The EPS09 package to be released soon, will contain both the NLO and LO results for the best fits and  the uncertainty sets.}

\FullConference{4th international workshop High-pT physics at LHC 09\\
                 February 4-7, 2009\\
                 Prague, Czech Republic}

\begin{document}
\section{Introduction}

The global analyses of the free-nucleon parton distribution functions (PDFs) are based on three fundamental aspects of QCD: asymptotic freedom, collinear factorization and scale evolution of the PDFs. These properties allow to compute the inclusive hard process cross-sections, schematically
$$
\sigma_{AB\rightarrow h+X} = \sum_{i,j} f_{i}^A(Q^2) \otimes \hat{\sigma}_{ij\rightarrow h+X} \otimes f_{j}^B(Q^2) + \mathcal{O}\left( {Q^{2}}\right)^{-n},
$$
where $f_{i}$s denote the universal, process-independent PDFs obeying the DGLAP \cite{DGLAP} evolution equations, and $\hat{\sigma}_{ij\rightarrow h+X}$ are perturbatively computable coefficients. This approach has proven to work extremely well with increasingly more different types of data included in the analyses. In the case of bound nucleons the validity of factorization is not as well established (see e.g. Ref. \cite{Armesto:2006ph}), but it has nevertheless turned out to provide a very good description of the world data from deep inelastic scattering (DIS) and Drell-Yan (DY) dilepton measurements involving nuclear targets \cite{Eskola:1998iy,Eskola:1998df,Eskola:2007my,Eskola:2008ca,Hirai:2007sx,deFlorian:2003qf}: in the nuclear environment the shape of the PDFs, however, is different from the free-nucleon PDFs. Here, we give an overview of our recent NLO global analysis of the nuclear PDFs (nPDFs) and their uncertainties \cite{Eskola:2009uj}.
 
\section{Analysis method and framework}

Our analysis follows the usual global DGLAP procedure:
\begin{itemize}
\item
{\bf The PDFs are parametrized at a chosen initial scale $Q_0^2$ imposing the sum rules.} In this work we do not parametrize the absolute nPDFs, but the nuclear modification factors $R_i^A(x,Q_0^2)$ encoding the relative difference to the free-proton PDFs through
\begin{equation}
f_{i}^A(x,Q^2) \equiv R_{i}^A(x,Q^2) f_{i}^{\rm CTEQ6.1M}(x,Q^2).
\label{eq:partondefinition}
\end{equation}  
Above, $f_{i}^{\rm CTEQ6.1M}(x,Q^2)$ refers to the CTEQ6.1M set of free proton PDFs \cite{Stump:2003yu} in the $\overline{MS}$ scheme and zero-mass variable flavour-number scheme. We consider three different modification factors: $R_V^A(x,Q_0^2)$ for both $u$ and $d$ valence quarks, $R_S^A(x,Q_0^2)$ for all sea quarks, and $R_G^A(x,Q_0^2)$ for gluons.
\item
{\bf The absolute PDFs at the parametrization scale $Q_0^2$ are connected to other perturbative scales $Q^2 > Q_0^2$ through the DGLAP evolution.} In this way, also the nuclear modification factors $R_i^A(x,Q^2)$ become scale-dependent and the initial flavour-independence may also disappear. An efficient numerical solver for the parton evolution is an indispensable tool for any global analysis, but in the case of nPDFs this is even more critical as we always need to perform the evolution separately for 13 different nuclei --- an order of magnitude more computation time than what is needed in a free-proton analysis.
\item
{\bf The cross-sections are computed using the factorization theorem.} In the present analysis the DIS and DY data constitute the main part of the experimental input, but we also employ the midrapdity $\pi^0$-production data measured in d+Au collisions at RHIC. Inclusion of the $\pi^0$-data provides important further constraints for the gluon modification that are partly complementary to the DIS and DY data.
\item
{\bf The computed cross-sections are compared with the experimental data, and the initial parametrization is varied until the best agreement with the data is found.} Our definition for the ``best agreement'' is based on minimizing the following $\chi^2$-function
\begin{eqnarray}
\chi^2(\{a\})   & \equiv & \sum _N w_N \, \chi^2_N(\{a\}) 
\label{eq:chi2mod_1}
\\	
\chi^2_N(\{a\}) & \equiv & \left( \frac{1-f_N}{\sigma_N^{\rm norm}} \right)^2 + \sum_{i \in N}
\left[\frac{ f_N D_i - T_i(\{a\})}{\sigma_i}\right]^2.
\label{eq:chi2}
\end{eqnarray}
Within each data set $N$, $D_i$ denotes the experimental data value with $\sigma_i$ point-to-point uncertainty, and $T_i$ is the theory prediction corresponding to a parameter set $\{a\}$. If an overall normalization uncertainty $\sigma_N^{\rm norm}$ is specified by the experiment, the normalization factor $f_N \in [1-\sigma_N^{\rm norm},1+\sigma_N^{\rm norm}]$ is introduced. Its value is determined by minimizing $\chi^2_N$ and the final $f_N$ is thus an output of the analysis. The weight factors $w_N$ are used to amplify the importance of those data sets whose content is physically relevant, but contribution to $\chi^2$ would otherwize be too small to have an effect on the automated $\chi^2$ minimization.
\end{itemize}

In addition to finding the parameter set $\{a^0\}$ that optimally fits the experimental data, quantifying the uncertainties stemming from the experimental errors has become an increasingly important topic in the context of global PDF analyses. The Hessian method \cite{Pumplin:2001ct} provides a practical way of treating this issue. It is based on a quadratic approximation for the $\chi^2$ around its minimum $\chi^2_0$,
\begin{equation}
\chi^2 \approx \chi^2_0 + \sum_{ij} \frac{1}{2} \frac{\partial^2 \chi^2}{\partial a_i \partial a_j} (a_i-a_i^0)(a_j-a_j^0) \equiv \chi^2_0 +  \sum_{ij} H_{ij}(a_i-a_i^0)(a_j-a_j^0),
\label{eq:chi_2approx}
\end{equation}
which defines the Hessian matrix $H$. Non-zero off-diagonal elements in the Hessian matrix are a sign of correlations between the original fit parameters, invalidating the standard (diagonal) error propagation formula
\begin{equation}
(\Delta X)^2 = \left( \frac{\partial X}{\partial a_1} \cdot \delta a_1 \right)^2 + \left( \frac{\partial X}{\partial a_2} \cdot \delta a_2 \right)^2 + \cdots
\end{equation}
for a PDF-dependent quantity (cross section) $X$. Therefore, it is useful to diagonalize the Hessian matrix, such that
\begin{equation}
\chi^2 \approx  \chi^2_0 +  \sum_{i} z_i^2,
\label{eq:chi_2diag}
\end{equation}
where each $z_i$ is a certain linear combination of the original parameters around $\{a^0\}$. In these variables, the usual form of the error propagation
\begin{equation}
(\Delta X)^2 = \left( \frac{\partial X}{\partial z_1} \cdot \delta z_1 \right)^2 + \left( \frac{\partial X}{\partial z_2} \cdot \delta z_2 \right)^2 + \cdots
\label{eq:error_better}
\end{equation}
stands on a much more solid ground. How to determine the size of the deviations $\delta z_k$ is, however, a difficult issue where no universally agreed procedure exists. We consider 15 parameters in the expansion (\ref{eq:chi_2diag}) and the prescription which we adopt here is to choose each $\delta z_k$ such that $\chi^2$ grows by a certain fixed amount $\Delta \chi^2$. Requiring each data set to remain close to its 90\%-confidence range, we end up with a choice $\Delta \chi^2=50$ (see Appendix A of Ref. \cite{Eskola:2009uj} for details). 

Much of the practicality of the Hessian method resides in constructing PDF error sets, which we denote by $S_k^\pm$. Each $S_k^\pm$ is obtained by displacing the fit parameters to the positive/negative direction along $z_k$ such that $\chi^2$ grows by the chosen $\Delta \chi^2=50$. Approximating the derivatives in Eq.~(\ref{eq:error_better}) by a finite difference, we may then re-write the error formula as
\begin{equation}
(\Delta X)^2 = \frac{1}{4} \sum_k \left[ X(S^+_k)-X(S^-_k) \right]^2,
\label{eq:error_best}
\end{equation}
where $X(S^\pm_k)$ denotes the value of the quantity $X$ computed by the set $S_k^\pm$. If the lower and upper uncertainties $\Delta X^\pm$ differ, they should be computed separately, using the prescription \cite{Nadolsky:2001yg}
\begin{eqnarray}
(\Delta X^+)^2 & \approx & \sum_k \left[ \max\left\{ X(S^+_k)-X(S^0), X(S^-_k)-X(S^0),0 \right\} \right]^2 \label{eq:ASymmetricPDFerrors} \\
(\Delta X^-)^2 & \approx & \sum_k \left[ \max\left\{ X(S^0)-X(S^+_k), X(S^0)-X(S^-_k),0 \right\} \right]^2, \nonumber
\end{eqnarray}
where $S^0$ denotes the best fit.

Along with the grids and interpolation routine for the best NLO and LO fits, our new release --- a computer routine called EPS09 \cite{EPS09code} --- contains also 30 error sets required for computation of the uncertainties to nuclear cross-section ratios similar to those we present in the following section. However, as can be understood from the definition (\ref{eq:partondefinition}), the total uncertainty in the absolute nPDFs is a combination of uncertainties from the baseline set CTEQ6.1M and those from the nuclear modifications $R_i^A$. Therefore, if an absolute cross-section is computed, also the CTEQ6.1M error-sets (with the EPS09 central set) should be included when calculating its uncertainty.

\section{Results}

\begin{figure}[!htb]
\center
\includegraphics[scale=0.5]{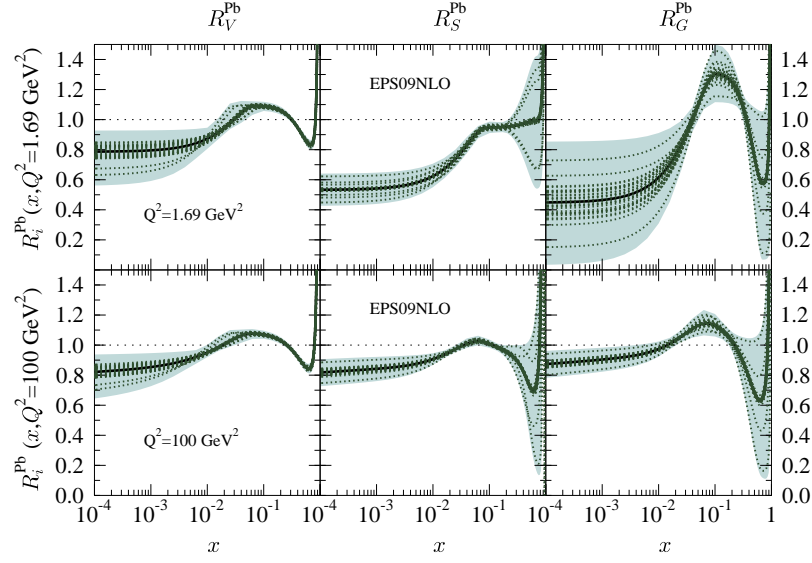}
\caption[]{\small The nuclear modifications $R_V$, $R_S$, $R_G$ for Lead at our initial scale $Q^2_0=1.69 \, {\rm GeV}^2$ and at $Q^2=100 \, {\rm GeV}^2$. The thick black lines indicate the best-fit results, whereas the dotted green curves denote the individual error sets. The shaded bands are computed using Eq.~(\ref{eq:ASymmetricPDFerrors}).}
\label{Fig:PbPDFs}
\end{figure}

In this section we present a selection of results from the NLO analysis. First, in Fig.~\ref{Fig:PbPDFs} we plot the obtained modifications at two scales, at $Q^2_0=1.69 \, {\rm GeV}^2$ and at $Q^2=100 \, {\rm GeV}^2$, which is to demonstrate the scale-dependence of the modifications and their uncertainties. One prominent feature to be noticed is that even if there is a rather large uncertainty band for the small-$x$ gluon modification $R_G^A$ at $Q^2_0$, the scale evolution tends to bring even a very strong gluon shadowing close to no shadowing at $Q^2\sim 100 \, {\rm GeV}^2$ --- a very clear prediction of the DGLAP approach. 

\begin{figure}[!htb]
\center
\includegraphics[scale=0.35]{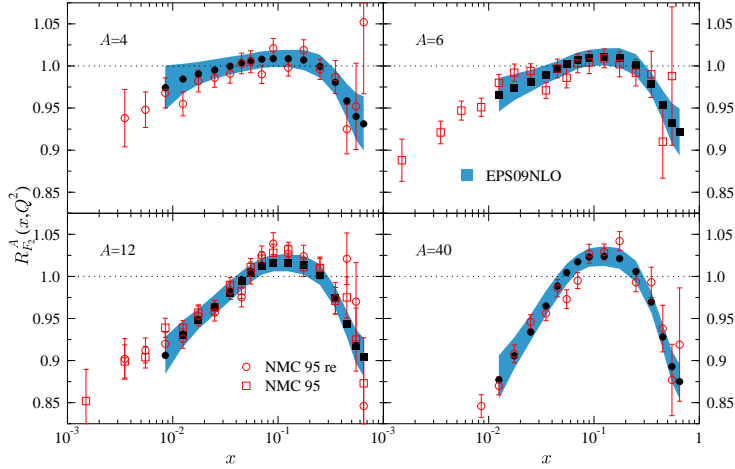}
\caption[]{\small The calculated NLO $R_{F_2}^A(x,Q^2)$ (filled symbols and error bands) compared with the NMC 95 \cite{Arneodo:1995cs} and the reanalysed NMC 95 \cite{Amaudruz:1995tq} data (open symbols).}
\label{Fig:RF2A1}
\end{figure}

\begin{figure}[htbp]
\center
\includegraphics[scale=0.40]{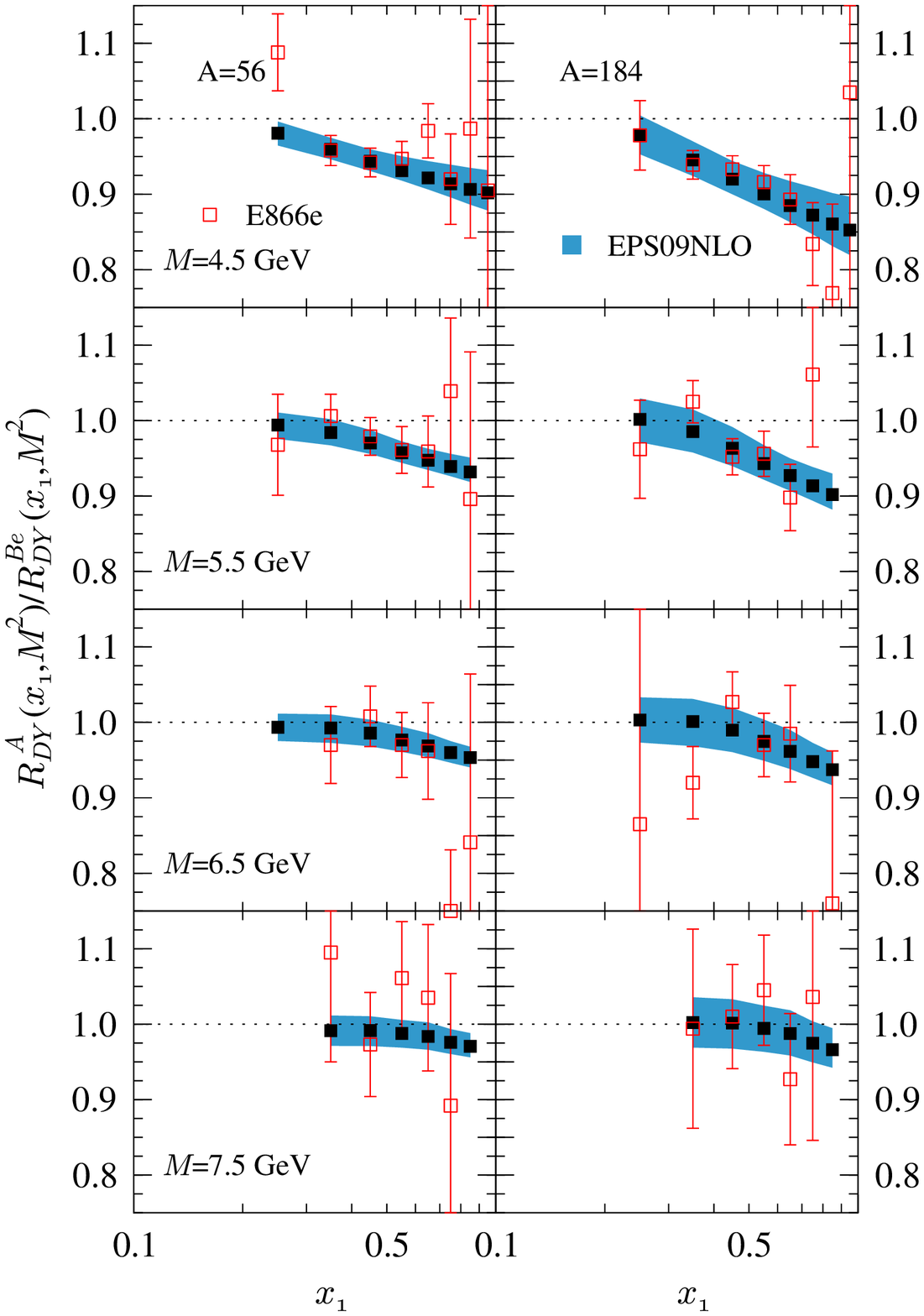}
\hspace{-1.0cm}
\includegraphics[scale=0.39]{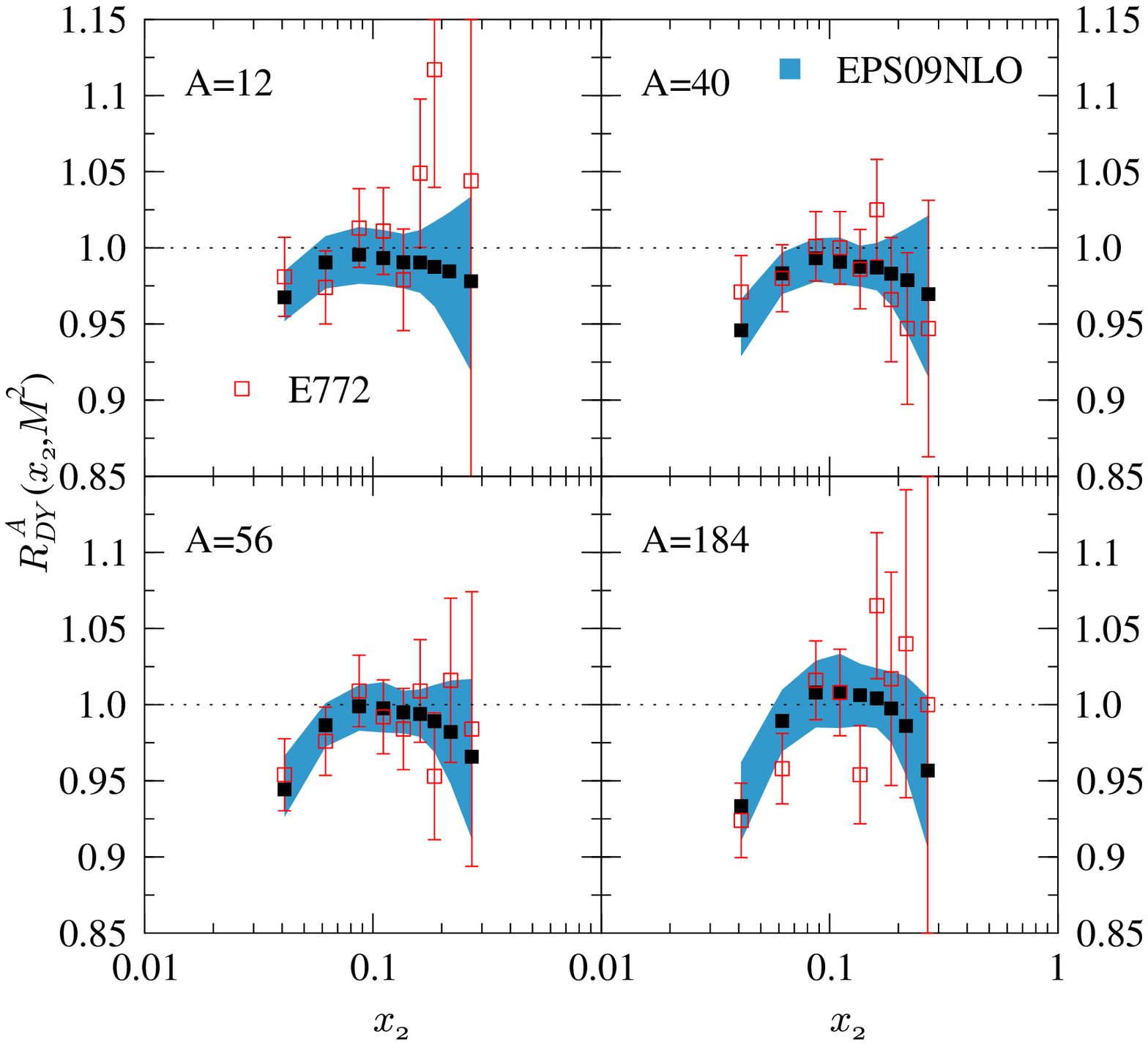}
\caption[]{\small The computed NLO $R_{\rm DY}^{\rm A}(x,M^2)$ (filled squares and error bands) as a function of $x_1$ and $x_2$ compared with the E866 \cite{Vasilev:1999fa} and the E772 \cite{Alde:1990im} data (open squares).}
\label{Fig:RDY}
\end{figure}

The DIS data constitutes the bulk of the experimental data available for the global analysis of nPDFs. In Fig.~\ref{Fig:RF2A1} we show some of the measured nuclear modifications for ($l+A$) DIS structure functions with respect to Deuterium,
\begin{equation}
R_{F_2}^{\rm A}(x,Q^2) \equiv  \frac{F_2^A(x,Q^2)}{F_2^d(x,Q^2)},
\end{equation}
and the comparison with our EPS09 NLO results.
The shaded bands denote the uncertainty derived from the 30 EPS09NLO error sets and, as should be noticed, their width is comparable to the error bars in the experimental data. This \emph{a posteriori} supports the validity of our procedure for determining the $\Delta \chi^2$. Similar conclusion can be drawn upon inspecting the nuclear effects in the Drell-Yan data
\begin{equation}
R_{\rm DY}^{\rm A}(x_{1,2},M^2) \equiv \frac{\frac{1}{A}d\sigma^{\rm pA}_{\rm 
DY}/dM^2dx_{1,2}}{\frac{1}{2}d\sigma^{\rm pd}_{\rm DY}/dM^2dx_{1,2}},
\end{equation}
where $M^2$ is the invariant mass of the lepton pair and $x_{1,2} \equiv \sqrt{M^2/s}\,e^{\pm y}$ ($y$ is the pair rapidity). Comparison to the E772 and E866 data is shown in Fig.~\ref{Fig:RDY}. Let us mention that the E772 data at $x_2 > 0.1$ carry some residual sensitivity also to the gluons and sea quarks, which has not been noticed before.
\begin{figure}[!h]
\centering
\includegraphics[width=18pc]{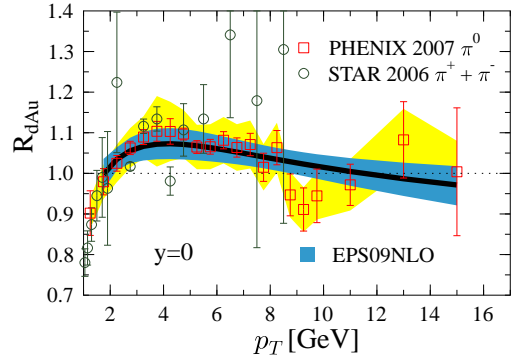}
\caption[]{\small 
{The computed $R_{\rm dAu}$ (thick black line and blue error band) at $y=0$} for inclusive pion production compared with the PHENIX \cite{Adler:2006wg} data (open squares). The error bars are the statistical uncertainties, and the yellow band indicates the point-to-point systematic errors. The PHENIX data have been multiplied by the optimized normalization factor $f_N = 1.03$, which is an output of our analysis. The STAR data \cite{Adams:2006nd} (open circles), multiplied by $f_N = 0.90$, are also shown although they were not included in the EPS09 analysis.}
\label{Fig:PHENIX}
\end{figure}

The nuclear modification for inclusive pion production in d+Au collisions relative to p+p is defined as 
\begin{equation}
R_{\rm dAu}^{\pi}  \equiv  \frac{1}{\langle N_{\rm coll}\rangle} \frac{d^2 N_{\pi}^{\rm dAu}/dp_T dy}{d^2 N_{\pi}^{\rm pp}/dp_T dy} \stackrel{\rm min. bias}{=} \frac{\frac{1}{2A} d^2\sigma_{\pi}^{\rm dAu}/dp_T dy}{d^2\sigma_{\pi}^{\rm pp}/dp_T dy}, \nonumber 
\end{equation}
where $p_T$, $y$ are the transverse momentum and rapidity of the pion, and 
$\langle N_{\rm coll}\rangle$ denotes the number of binary nucleon-nucleon collisions. A comparison with the PHENIX and STAR data is shown in Fig.~\ref{Fig:PHENIX}, and evidently, the shape of the spectrum gets very well reproduced by our parametrization. The downward trend toward large $p_T$ provides direct evidence for an EMC-effect in the large-$x$ gluons, while the suppression toward small $p_T$ is in line with the gluon shadowing. No additional effects are needed to reproduce the observed spectra. It is also reassuring that this shape is practically independent of the fragmentation functions used in the calculation --- modern sets like \cite{Kniehl:2000fe,Albino:2008fy,deFlorian:2007aj} all give equal results.

\begin{figure}[!htb]
\center
\includegraphics[scale=0.40]{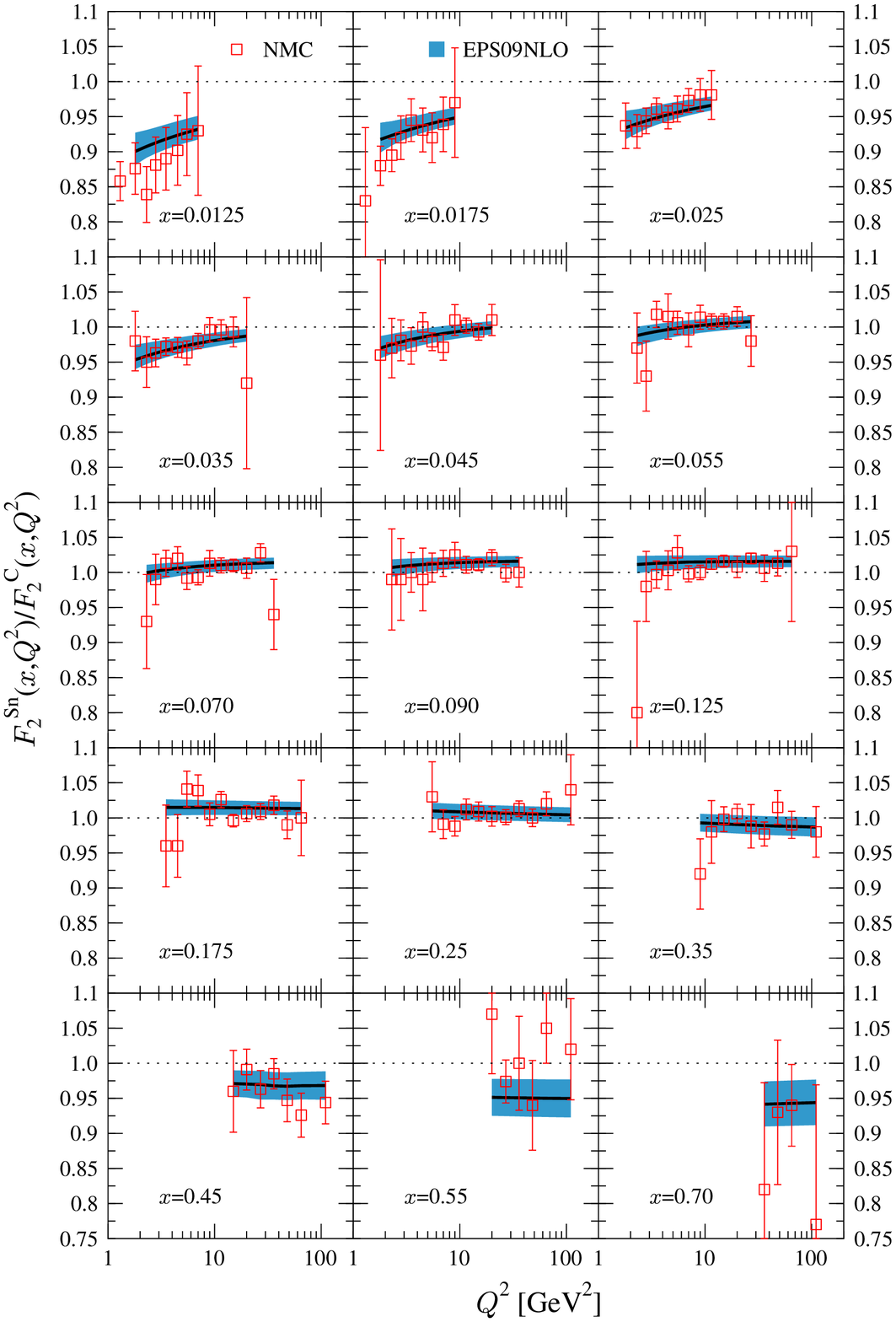}
\includegraphics[scale=0.34]{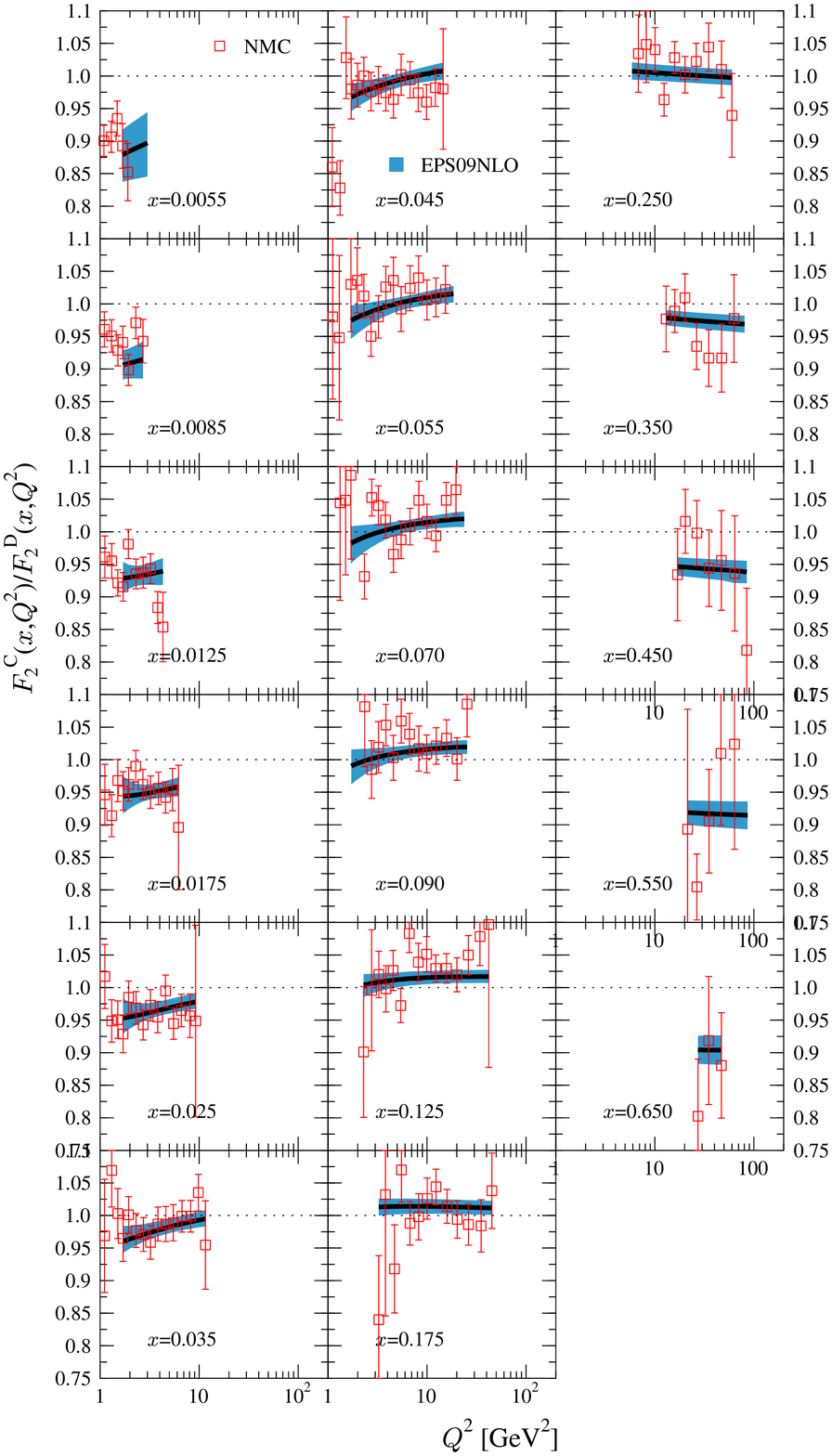}
\caption[]{\small The calculated NLO scale evolution (solid black lines and error bands) of the ratio $F_2^{\mathrm{Sn}}/F_2^{\mathrm{C}}$ and $F_2^{\mathrm{C}}/F_2^{\mathrm{D}}$, compared with the NMC data \cite{Arneodo:1996ru} for  several fixed values of $x$.}
\label{Fig:RF2_slopes}
\end{figure}

Special attention should be paid to the scale-breaking effects in the data and to their good description by the DGLAP evolution. These effects are clearly visible e.g. in the E886 Drell-Yan data in Fig.~\ref{Fig:RDY} where the trend of diminishing nuclear effects toward larger invariant mass $M^2$ is observed. Also, from the DIS data as a function of $Q^2$, shown in Fig.~\ref{Fig:RF2_slopes}, the general features can be filtered out: At small $x$ the $Q^2$-slopes tend to be positive, while toward larger $x$ the slopes gradually die out and become even slightly negative.

\begin{figure}[!htb]
\center
\includegraphics[scale=0.45]{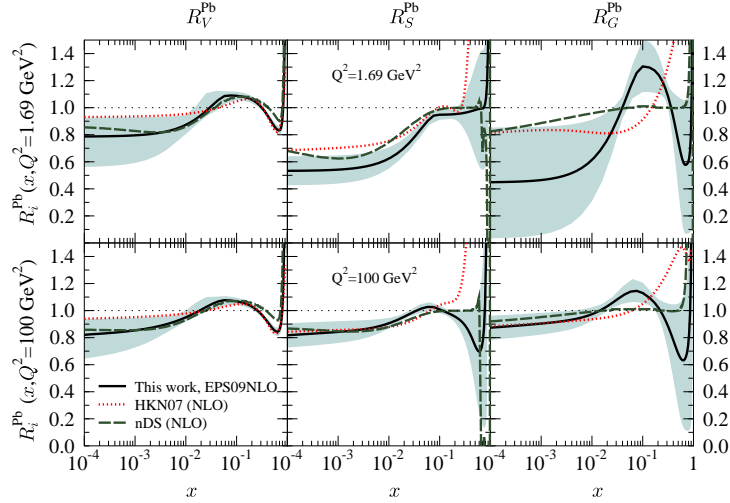}
\caption[]{\small Comparison of the nuclear modifications for Lead at $Q^2 = 1.69 \, {\rm GeV}^2$ and at $Q^2 = 100 \, {\rm GeV}^2$ from the NLO global DGLAP analyses HKN07~\cite{Hirai:2007sx}, nDS~\cite{deFlorian:2003qf} and this work, EPS09.}
\label{Fig:NLOcomp}
\end{figure}

To close this section we show, in Fig.~\ref{Fig:NLOcomp}, a comparison of the nuclear modifications for Lead from all available global NLO analyses. The comparison is shown again at $Q^2_0=1.69 \, {\rm GeV}^2$ and at $Q^2=100 \, {\rm GeV}^2$. Most significant differences --- that is, curves being outside our error bands --- are found from the sea quarks and gluons. At low $x$, the differences shrink when the scale $Q^2$ is increased, but at the high-$x$ region notable discrepancies persist at all scales. Most of the differences are presumably explainable by the assumed behaviours of the fit functions, but also the choices for the considered data sets (e.g. HKN and nDS do not implement the pion data), and differences in the definition of $\chi^2$ carry some importance.

\section{Hadrons in the forward direction in d+Au at RHIC?}

\begin{figure}[htbp]
\center
\includegraphics[scale=0.45]{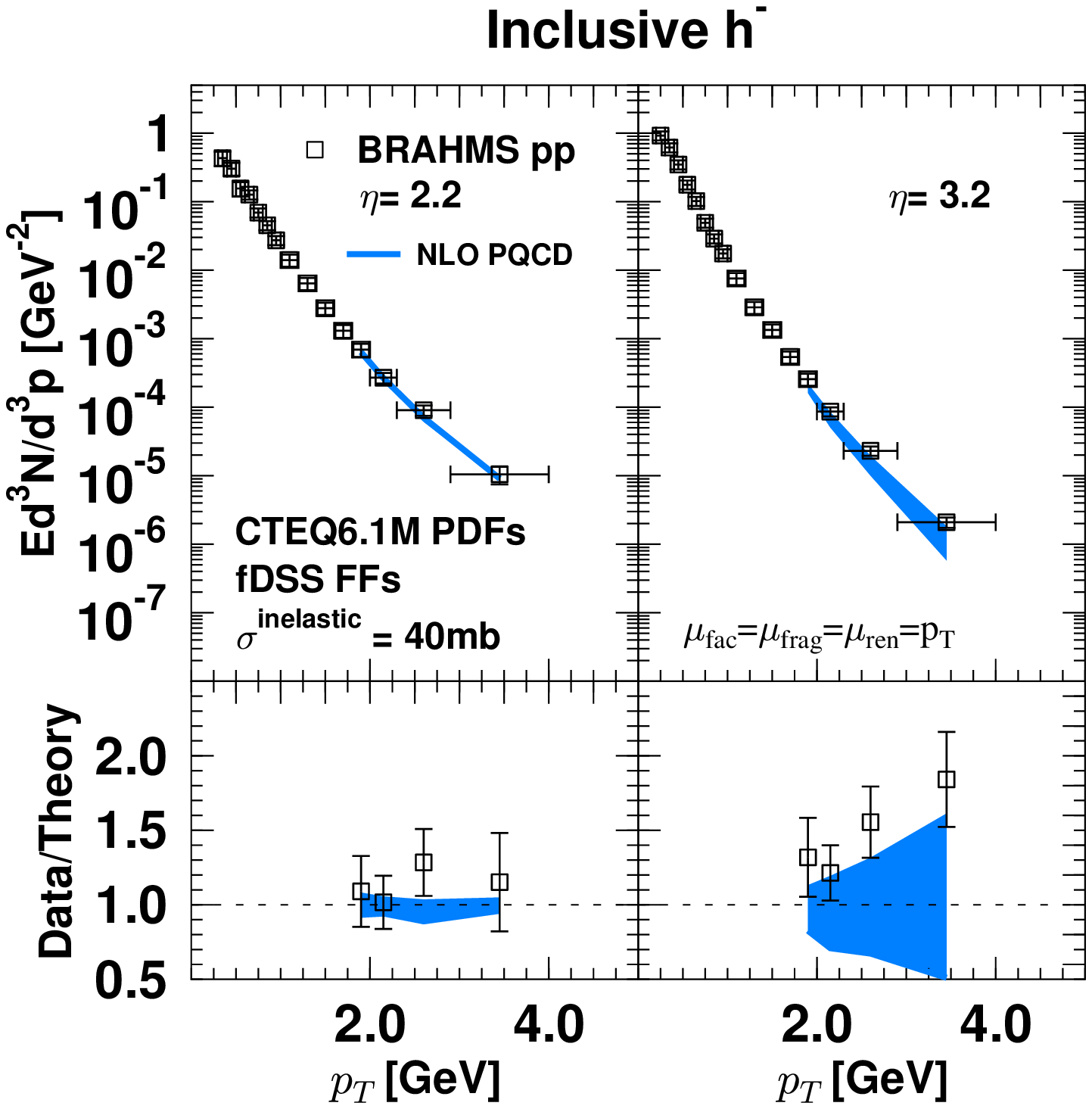}
\includegraphics[scale=0.45]{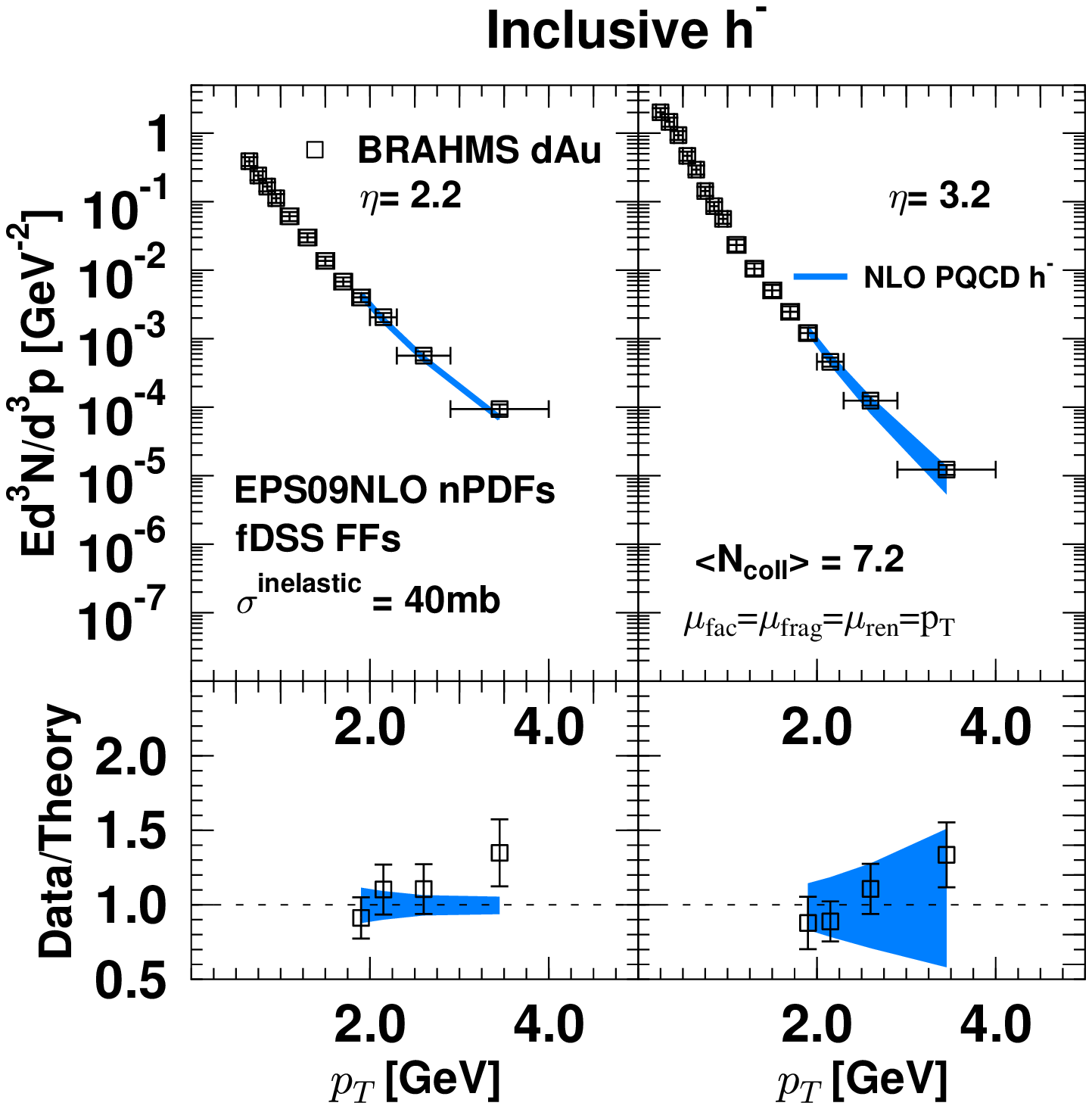}
\caption[]{\small Inclusive $h^-$ yield in p+p and d+Au collisions. The experimental $p_T$-binned data \cite{Arsene:2004ux} are shown by open squares with statistical and systematical errors added in quadrature. The blue band indicates the 90\% confidence range derived from the CTEQ6.1M and EPS09 uncertainties. The calculated cross-sections have been averaged over the $p_T$-bin width.}
\label{Fig:BRAHMS1}
\end{figure}

In our previous article \cite{Eskola:2008ca}, we studied the case of a very strong gluon shadowing, motivated by the suppression in the nuclear modification $R_{\rm dAu}$ for the negatively-charged hadron yield in the forward rapidities ($\eta=2.2,3.2$) measured by the BRAHMS collaboration \cite{Arsene:2004ux} in d+Au collisions at RHIC. We now return to this issue by applying the EPS09NLO parametrization to this specific process. The EPS09 predictions (with fDSS fragmentation functions \cite{de Florian:2007hc}) compared with the BRAHMS data for the absolute $h^-$ spectra,
\begin{equation}
\frac{d^3 N^{\rm pp}}{d^2p_T dy} \stackrel{\rm min. bias}{=} \frac{1}{\sigma_{NN}^{\rm inelastic}} \frac{d^3\sigma^{\rm pp}}{d^2p_T dy}
\quad ; \quad
\frac{d^3 N^{\rm dAu}}{d^2p_T dy} \stackrel{\rm min. bias}{=} \frac{{\langle N_{\rm coll}\rangle}}{\sigma_{NN}^{\rm inelastic}} \frac{\frac{1}{2A}d^3\sigma^{\rm dAu}}{d^2p_T dy},
\label{eq:BRAHMScrosssection}
\end{equation}
are shown in Fig.~\ref{Fig:BRAHMS1}. In the $\eta=2.2$ bin, the measured p+p and d+Au spectra are both in good agreement with the NLO pQCD. However, in the most forward $\eta=3.2$ bin in the p+p case, there is a systematic and significant discrepancy between the measured and computed $p_T$ spectra. This observation casts doubts on any conclusion made from the nuclear modification $R_{\rm dAu}$ alone --- clearly, one should first account for the absolute baseline spectrum in p+p collisions. This is why we have not included this BRAHMS data set in our global analysis, either.

\section{Summary}

We have here outlined our NLO analysis of nPDFs. The very good agreement with the experimental data $\chi^2/N \approx 0.79$ --- especially the correct description of the scaling-violation effects --- lends support to the validity of collinear factorization in nuclear collisions. In addition to the best fit, we have distilled the experimental errors into the 30 nPDF error-sets, which encode the relevant parameter-space neighbourhood of the $\chi^2$ minimum. All these sets will be available as a computer routine at \cite{EPS09code} for general use. 

Although not discussed here, we have also performed the leading-order (LO) counterpart of the NLO analysis as we want to provide the uncertainty tools also for the widely-used LO framework. The best-fit quality is very similar both in LO and NLO, but the uncertainty bands become somewhat smaller when going to higher order. In the LO case, our error analysis indeed accommodates also the strong gluon shadowing suggested in our previous work EPS08 \cite{Eskola:2008ca}. 

In the near future, more RHIC data will become available and published, and factorization will be tested further.
Also p+$A$ runs at the LHC would be welcome for this purpose. Even better possibilities for the nPDF studies would, however, be provided by the lepton-ion colliders like the planned eRHIC or LHeC.

During the recent years, the free-proton PDF analyses have gradually shifted to a better organized prescription for treating the heavy quarks than the zero-mass scheme employed in this analysis and in CTEQ6.1M. Such general-mass scheme should be especially important e.g. in the case of charged-current neutrino interactions. Interestingly, the CTEQ collaboration has lately looked also at this type of data among other neutrino-Iron measurements \cite{Schienbein:2007fs}, and noticed that the best fit tends to point to somewhat different nuclear modifications than what have been obtained in the global nPDF analyses. Both the extension of the nPDF analysis to the general-mass scheme and a systematic investigation of the possible discrepancies between neutrino and other data, remains as future work.

\end{document}